\documentstyle[12pt,aaspp4,tighten]{article} 

\def\pagebreak{\vfill\eject}

\def\Msun{{M$_\odot$}}
\def\sun{\mbox{$_\odot$}}

\def\deg{{$^\circ$}}
\def\half{{\leavevmode\kern.1em\raise.5ex
\hbox{\the\scriptfont0 1}\kern-.1em /
\kern-.15em\lower.25ex\hbox{\the\scriptfont02}}} 
\def\gtsim{\lower.5ex\hbox{$\buildrel > \over\sim$}}
\def\ltsim{\lower.5ex\hbox{$\buildrel < \over\sim$}}
\def\sun{\mbox{$_\odot$}}
\def\kms{~km~s$^{-1}$}

\def\ceo{C$^{18}$O}
\def\ceof{C$^{18}$O(J=1$-$0)}

\def\h{H$_2$}

\def\water{H$_2$O}

\def\nht{NH$_3$}

\def\jyb{~Jy~beam$^{-1}$}

\def\mjybkms{~mJy~beam$^{-1}$ km~s$^{-1}$}
\def\mjyb{~mJy~beam$^{-1}$}
\def\ra#1#2#3{#1$^{\rm h}$#2$^{\rm m}$#3$^{\rm s}$}
\def\dec#1#2#3{$#1^\circ#2'#3''$}

\def\HII{\mbox{{\rm H}\,{\scriptsize II}}}

\slugcomment{ApJ, to be published 10 October 2004, v614 issue}

\begin{document}

\title{The Circumstellar Environment of the Early B Protostar
  G192.16--3.84 and the Discovery of a Low-Mass, Protostellar Core} 

\author{D. S. Shepherd\altaffilmark{1}, 
  T. Borders\altaffilmark{2},
  M. Claussen\altaffilmark{1},
  Y. Shirley\altaffilmark{1},  \&
  S. Kurtz\altaffilmark{3}}

\vspace{-3mm}
\altaffiltext{1}{National Radio Astronomy Observatory, P.O. Box 0,
Socorro, NM 87801}
\altaffiltext{2}{Sonoma State University, Rohnert Park, CA 94928-3609}
\altaffiltext{3}{Centro de Radioastronom\'\i a y Astrof\'\i sica,
Universidad Nacional Aut\'onoma de M\'exico, Apdo. Postal 3-72, C.P.
58089, Morelia, Mich. Mexico}

\begin{abstract}

\vspace{4mm}

We have observed the massive star forming region associated with the
early B protostar G192.16--3.84 in {\nht}(1,1), 22.2\,GHz {\water}
masers, 1.3\,cm continuum emission, and at 850\,$\mu$m.  The dense gas
associated with G192.16 is clumpy, optically thin, and has a mass of
0.9\,{\Msun}.  The {\nht} core is gravitationally unstable which may
signal that the outflow phase of this system is coming to an end.
Water masers trace an ionized jet $0.8''$ (1600 AU at a distance of
2\,kpc) north of G192.16.  Masers are also located within 500\,AU of
G192.16, their velocity distribution is consistent with but does not
strongly support the interpretation that the maser emission arises in
a 1000\,AU rotating disk centered on G192.16.  Roughly $30''$ south of
G192.16 (0.3\,pc) is a compact, optically thick ($\tau = 1.2$) {\nht}
core (called G192\,S3) with an estimated mass of 2.6\,{\Msun}.  Based
on the presence of 850\,$\mu$m and 1.2\,mm continuum emission,
G192\,S3 probably harbors a very young, low-mass protostar or
proto-cluster.  The dense gas in the G192\,S3 core is likely to be
gravitationally bound and may represent the next site of star
formation in this region.
  
\end{abstract}


\section{Introduction}

The wealth of observational evidence suggests that outflows from
low-mass protostars are almost certainly powered by disk accretion
where magnetic stresses control the balance between inflow and outflow
and determine the outflow structure far from the protostar.
Intermediate mass protostars, up to early B stars, can probably also
be described by this general picture however the details are not as
clear.  The presence of accretion disks and outflows traced to within
50--100~AU of early B protostars suggests that they are also likely
formed via accretion (e.g., Shepherd 2003, Chini et al. 2004).

The structure and kinematics of circumstellar material around early B
protostars directly affects the formation and evolution of the central
object(s).  Circumstellar disks provide a reservoir of material that
can transport angular momentum away from the central object in two
possible ways: 1) disks create outflows which transport angular
momentum (e.g., Shu et al. 2000; Ouyed \& Pudritz 1999); and 2) if
$M_{disk}/M_\star > 0.3$ the disk may be locally unstable and provide
an additional means of transporting angular momentum away from the
protostar (e.g., Laughlin \& Bodenheimer 1994; Yorke, Bodenheimer, \&
Laughlin 1995; Shepherd, Claussen, \& Kurtz 2001, hereafter SCK01).

Massive star formation tends to occur in compact clusters and it can
be difficult to separate individual outflows from each other (e.g.,
Beuther, Schilke, \& Stanke 2003; Beuther \& Schilke 2004).  Even when
a single, early B protostar dominates the flow dynamics in a cluster,
companions or close members of a cluster can interact with an
accretion disk causing it to precess or be truncated (e.g., Fendt \&
Zinnecker 1998 and references therein; Shepherd et al. 2000).  The
strong radiation field from massive protostars and cluster members can
also significantly affect the cluster and disk evolution (e.g.,
K\"onigl 1999; Yorke \& Sonnhalter 2002).

Understanding the interactions between cluster members, dense gas, and
the radiation field is important to be able to unravel the processes
that affect the formation and evolution of massive star forming
regions.  One early B protostar that may shed some light on this issue
is G192.16--3.84 (hereafter G192.16, also known as IRAS 05553+1631).
G192.16 is the most massive object in a protostellar cluster at a
distance of 2\,kpc.  An ultracompact (UC) \HII\ region with luminosity
$\sim 3 \times 10^{3}$ {{L$_\odot$}} implies the presence of an early
B star with a mass of 8 to 10\,M{\sun} (Hughes \& MacLeod 1993;
SCK01).  The detection of {\nht}(1,1) (Molinari et al. 1996), strong
22 GHz {\water} maser emission (Codella, Felli, \& Natale 1996), and 1
\& 3\,mm continuum emission (Shepherd et al. 1998; Beuther et
al. 2002) also suggests that this is a site of recent star formation.
G192.16 is at the center of a 100\,M{\sun} molecular outflow with mass
outflow rate $\dot{M}_{flow} \sim 10^{-4}$\,M{\sun}\,yr$^{-1}$ and
dynamical age $\sim 2 \times 10^{5}$ years (Snell, Dickman, \& Huang
1990; Shepherd et al. 1998, Devine et al. 1999).  Recent 2.1\,$\mu$m
observations detect a very red object near the position of the
radio-detected UC\,\HII\ region.  If the 2.1\,$\mu$m source is a
direct detection of G192.16, then the protostellar photosphere is
extincted by $A_V = 29-33$ magnitudes (Indebetouw et al. 2003).

Based on 22.2\,GHz {\water} maser and centimeter continuum
observations, Shepherd \& Kurtz (1999) proposed that the G192.16
massive young stellar object (YSO) is surrounded by a 1000\,AU
rotating, molecular torus.  The size of this molecular torus or
possible disk is smaller than the handful of structures detected
around other early B protostars.  For example, IRAS\,20126+4104
appears to have a 1700\,AU flattened, rotating torus detected in
CH$_3$CN (Cesaroni et al. 1999) and a similar sized structure is
detected in {\nht} emission around AFGL 5142 (Zhang et al. 2002).  It
is unclear whether the circumstellar material detected in these dense
core tracers are flattened, rotating cores, tori, or large disks.
Significantly larger circumstellar structures are also observed: a
protostar in M17 appears to have a torus or flared disk of diameter
20,000\,AU as mapped with near-infrared imaging (Chini et al. 2004);
and flattened, rotating molecular cores are inferred from high density
tracers such as H$^{13}$CN (NGC 7538S; Sandell, Wright, \& Forster
2003) and {\nht} (IRAS\,20126+4104; Zhang, Hunter \& Sridharan 1998).

Approximately 1'' north of G192.16, {\water} masers are also detected
along a strip of ionized gas with a negative spectral index implying
the presence of a mixed synchrotron and thermal emission jet (Shepherd
\& Kurtz 1999).  The driving source for this collimated jet has not
been detected.

Continuum observations at 7 mm by SCK01 with a resolution of $\sim
40$\,mas did not detect the 1000\,AU torus predicted by Shepherd \&
Kurtz (1999).  Instead, compact warm dust emission was detected within
50\,mas of the massive YSO.  A model fit to the 7\,mm continuum
emission suggested an 8 {\Msun} central protostar (G192\,S1) with a
possible protostellar companion (G192\,S2), a 130 AU diameter disk
with a mass of 3-20 {\Msun}, and an outflow with $\sim 50$ degree
opening angle.  The non-detection of emission from the the 1000\,AU
torus by SCK01 could be the result of resolving out the torus with the
7mm observations or the dust temperature could be too low to emit
significantly in 7\,mm continuum.  Recent observations at 450 and
850\,$\mu$m (at resolution $8''$ and $14''$, respectively) show that
there is extended warm dust emission centered on G192.16 (Williams,
Fuller, \& Sridharan 2004).  The far-infrared emission is centrally
peaked and corresponds to a gas+dust mass of roughly 26 to 64{\Msun}
(Williams et al. assumed a kinematic distance of 2.5\,kpc; at a
distance of 2\,kpc the mass would be 16 to 40{\Msun}).

In an effort to understand the cluster and circumstellar environment
of G192.16, we present observations of {\nht}(1,1) tracing dense gas
and {\water} masers near the massive protostar, along with archive
data of 850\,$\mu$m continuum emission.

\section{Observations}

\subsection{Very Large Array} 

Observations of {\nht}(1,1) line emission at 23.694\,GHz were made
with the National Radio Astronomy Observatory's\footnote{The National
Radio Astronomy Observatory is a facility of the National Science
Foundation operated under cooperative agreement by Associated
Universities, Inc.}  Very Large Array (VLA) in ``D'' configuration.
Baselines between 35~m and 1.03~km could detect a largest angular
emission scale, $\theta_{LAS} \sim 60''$.  The estimated uncertainty
of the flux calibration is $\sim 3$\%.  Calibration and imaging was
performed using the AIPS$++$ data reduction package.  The continuum
emission was subtracted in the $uv$ plane.  The line-only emission was
imaged using Briggs robust $uv$ weighting and deconvolved with CLEAN.
Continuum-only emission was recovered and imaged separately using
natural $uv$ weighting and CLEAN deconvolution.

Observations of {\water} masers at 22.235\,GHz were made with the VLA
in ``B'' configuration. Calibration and
imaging was performed using the AIPS data reduction package.  The data
were imaged using natural $uv$ weighting and CLEAN deconvolution.  A
summary of the VLA observations is given in Table 1.

\small
\begin{table}[h!]
\caption{VLA Observation Summary}
\begin{tabular}{|l|c|c|c|}
\hline
  & {\nht}(1,1)	        & 1.3\,cm continuum$^\dagger$   & {\water} maser\\
\hline
\hline
Observation Date	
  &2003 April 25	& {\nodata}           &2001 June 17\\
Configuration		
  & D array		& {\nodata}           &B array\\
Rest Frequency (GHz)	
  &23.6945 		& {\nodata}           &22.2351\\
Beam Size (arcsec)
  &$2.93 \times 2.55$	&$2.92 \times 2.55$   &$0.68 \times 0.27$\\
Beam P. A. (degrees) 	
  &44.8			&44.8                 &--61.65\\
Resolution (\kms)	
  &0.31		        & {\nodata}	      &0.33\\
Total Bandwidth (\kms)  
  &39.4			& {\nodata}           &84.3\\
Gain Calibrator		
  &J0530+135		& {\nodata}	      &J0530+135\\
Flux Calibrator		
  &J0137+331 (3C\,48)	& {\nodata}	      &J0137+331 (3C\,48)\\
Number of Channels	
  &128			& 1                   &256\\
Peak Flux (\mjyb)	
  &25.3	                & 2.8	              &7,460.\\
Map RMS	(\mjyb)         
  &2.2			& 0.39                &28\\
\hline
\end{tabular}

\vspace{2mm}
$^\dagger$ Continuum emission subtracted from {\nht} line in the $uv$
plane.  Observational parameters are the same as for {\nht}. 

\end{table}
\normalsize

\subsection{Very Long Baseline Array}

Observations of {\water} masers were made at three epochs with a 2--4
month time separation using the Very Long Baseline Array (VLBA).
Observations were not phase-referenced, thus absolute position
information was lost.  Calibration and imaging was performed using the
AIPS data reduction package.  The data were imaged using natural $uv$
weighting and the image was deconvolved with CLEAN.  Masers were
detected in each observation and were quite variable with flux
densities ranging from 15{\mjyb} to 230{\jyb}.  A summary of the VLBA
observations is given in Table 2.

In all three VLBA epochs, masers were present at two locations in the
image.  The epochs were aligned assuming the brightest maser spot
remained constant in position.  Based on the scatter distribution of
maser features, the relative positions between epochs are probably
accurate to within 5\,mas.  Because absolute position information was
not available, proper motions of individual masers could not be
determined.

An estimate of the absolute position of the VLBA masers was obtained
by comparing the brightest VLBA {\water} masers observed on 2001 May
01 with the brightest VLA masers observed on 2001 June 17 at similar
velocities.  Although the VLA and VLBA observations were not obtained
simultaneously, the positions of three masers with relative positions
(270\,mas, -495\,mas), (290\,mas, 175\,mas), and (0\,mas, 0\,mas)
remained roughly constant.  The three spots from each observation were
aligned to within 5\,mas.  Thus, the relative error between the VLA
and all VLBA images is 7\,mas (adding positional errors in quadrature)
while the absolute error of the VLA image is $\sim 0.1''$.

\small
\begin{table}[h]
\caption{VLBA Observation Summary}
\begin{tabular}{|l|c|c|c|}
\hline
			&Epoch 1	&Epoch 2	&Epoch 3\\
\hline
\hline
Observation Date	&~~~~2001 May 04~~~~	
			&~~~~2001 July 02~~~~	
			&2001 November 17\\
Rest Frequency (GHz)	&22.2351	&22.2351	&22.2351 \\
Beam Size (mas)		&$0.86 \times 0.36$   	
			&$1.10 \times 0.33$    
			&$0.96 \times 0.36$ \\
Beam P. A. (degrees) 	&--5.30		&--15.35	&--13.09\\
Resolution (\kms)	&0.21	        &0.21	        &0.21\\
Total Bandwidth (\kms)~~ &53.9 	        &53.9		&53.9	\\
Bandpass Calibrator	&J0555+3948	&J0555+3948	&J0555+3948\\
Number of Channels	&256		&256		&256\\
Peak Flux (\jyb)	&8.7		&0.5 		&228.7\\
Map RMS	(\jyb)          &0.007	        &0.002	        &0.002\\
\hline
\end{tabular}
\end{table}

\subsection{James Clerk Maxwell Telescope}

Submillimeter continuum observations were obtained from the Canadian
Astronomy Data Centre Archive of SCUBA (Submillimeter Common User
Bolometer Array, Holland et al. 1999).  Observations from Williams et
al. (2004) on the night of 2000 March 18, were combined with earlier
observations from the night of 1997 October 18 to increase the
sensitivity in the final image.  Three 850\,$\mu$m jiggle maps were
reduced using the SURF (SCUBA User Reduction Facility) reduction
package (Jenness \& Lightfoot 1998).  Each map was flat-fielded and
corrected for chop throw (120$''$), extinction, and sky noise.  The
extinction was determined by using the $\tau_{225}$ opacities
determined at $225$ GHz with the tipper located at the Caltech
Submillimeter Observatory and scaling this value to the 850\,$\mu$m
optical depth, $\tau_{850}$, using the relationship derived by
Archibald et al. (2002).  The final 850\,$\mu$m image was made by
adding together the individual images with pixels sampled at the
Nyquist limit (7$''$).  The final image has a resolution of $15''$.
The image was calibrated using Uranus jiggle maps observed within a
few nights of G192.16.  The total estimated flux uncertainty is
$\sim$30\%\ based on comparisons of the peak flux in each separately
calibrated image.

\section{Results}

Figures 1 \& 2 present an overview of the interferometric observations
of {\nht}(1,1), 1.3\,cm continuum emission, and {\water} masers
observed with the VLA and VLBA.  G192.16, detected in 1.3\,cm
continuum emission at \ra{05}{58}{13.53}, \dec{16}{31}{58.3} (J2000),
is located within a region of clumpy {\nht} emission.  {\water} masers
are only detected within $1''$ of the massive protostar.  A compact
{\nht} core, G192 S3\footnote{SCK01 designated G192\,S1 as the B2
protostar and G192\,S2 as the proposed companion located $\sim 80$\,AU
north of G192\,S1.  Here, we use the term G192.16 to refer to the
S1+S2 system.}, is discovered $\sim 30''$ south of G192.16 at
\ra{05}{58}{13.67}, \dec{16}{31}{31.8} (J2000).  No centimeter
continuum emission or {\water} masers are detected near G192\,S3.
An image at 850\,$\mu$m shows a strong central peak at the position of
G192.16 along with an extension toward G192\,S3 (Fig. 3).

\subsection{Dense molecular gas and embedded protostars}

{\nht} emission near G192.16 is clumpy and covers a region of roughly
$20'' \times 10''$ ($\sim 0.2 \times 0.1$\,pc at a distance of
2\,kpc).  Figure 1 shows that the G192.16 protostar is located near
the center of the clumps.  The emission is detected in each channel at
a level of only $2 - 3 \sigma$ (Fig. 4).  Spectra averaged over 
G192.16 and G192\,S3 (Fig. 5) show that the G192.16 {\nht} emission
peaks at $v_{LSR} = 5.7${\kms} while G192\,S3 is shifted by 1{\kms}
to $6.7${\kms}.  The spectra show both the main inversion line and the
first two hyperfine components of the {\nht}(1,1) lines, although the
satellite lines toward G192.16 are detected at a level of only $3
\sigma$.  Gaussian fits to the main lines result in a peak of 4{\mjyb}
and FWHM of 1.52{\kms} for G192.16 and 13{\mjyb} and FWHM of
0.77{\kms} for G192\,S3.  Figure 6 shows a first moment map of the
intensity-weighted velocity field of the {\nht} emission,
$\left[\int I(v)v\,dv\right]/\left[\int I(v)dv\right]$.
There is no obvious velocity gradient in either region although the
narrow line width could easily prevent detection of a velocity
gradient in G192\,S3.   

The relative intensities of the main and satellite components of the
(1,1) transition can be used to estimate the optical depth of the main
component, $\tau$, from the equation (Ho \& Townes 1983):
\begin{equation}
\frac{\Delta T_A^*(J,K,m)}{\Delta T_A^*(J,K,s)}
 = \frac{1-e^{-\tau(J,K,m)}}{1-e^{-a\tau(J,K,m)}}
\end{equation}
where $T_A^*(J,K,m)$ is the measured antenna temperature of the (J,K)
transition for the main line, $s$ refers to the satellite line, and
$a$ is the satellite to main intensity (0.28 \& 0.22 for the (1,1)
transition).  The ratio of intensity of the (1,1) and (2,2)
transitions is sensitive to level populations and thus acts as a
temperature probe for cold molecular gas.  Unfortunately, we were not
able to observe the (2,2) transition.  Temperatures of cores in Orion
range from 15 to 40\,K with 30\,K being typical of warm cores in the
Orion {\nht} filaments (Wiseman \& Ho 1998).  Thus, we assume a
temperature of 30\,K.  Following the procedures and assumptions
described in Ho \& Townes (1983) and Wiseman \& Ho (1998) and assuming
an [{\nht}]/[{\h}] abundance ratio of $3 \times 10^{-8}$ (Harju,
Walmsley, \& Wouterloot 1993), we estimate the {\nht} column density,
{\h} number density, and total gas mass of the {\nht} cores (Table 3).
The last column in Table 3 gives the range of values found in {\nht}
cores in Orion.  Within the large uncertainties of our assumptions,
the clump properties in the G192.16 star forming region are similar to
those found in Orion.

\begin{table}[h!]
\caption{Properties of the ammonia cores}
\small
\begin{tabular}{|l|c|c||c|}
\hline
 & & & Comparison with \\
		&G192.16
		&G192\,S3
		&{\nht} Cores in Orion$^{\dagger}$\\

\hline
\hline
$\tau$ ({\nht} main component)		
                &0.06          		 
		&1.2
		&0.1-2.1 \\
{\nht} Column Density (cm$^{-2}$)	
		&~~~$8.1 \times 10^{13}$~~~
		&~~~$8.2 \times 10^{14}$~~~
		&~~~$1.9 \times 10^{12}~-~6.3 \times 10^{14}$\\ 
{\h} Number Density (cm$^{-3}$)    
		&$8.3 \times 10^{3}$
		&$1.6 \times 10^{5}$
		&$5.0 \times 10^{3}~-~6.3 \times 10^{5}$\\
Total Gas Mass (\Msun)	
		&0.9			
		&2.6
		&0.1-1.9\\  	   
\hline
\end{tabular}

\vspace{2mm}
$\dagger$~~Wiseman \& Ho (1998)
\end{table}
\normalsize

Ionized gas is detected in free-free continuum emission at 1.3\,cm at
the position of G192.16.  The source is unresolved with a flux density
of 2.8\,mJy ($7 \sigma$).  The number of Lyman continuum photons
required to produce the central peak is log\,$N_L =
44.87$\,photons~s$^{-1}$, which corresponds to a single
zero-age-main-sequence B2 star with $L_{bol} \sim 2.8 \times
10^3$\,L{\sun} (Thompson 1984).  This is consistent with the
luminosity derived by Shepherd et al. (1998) and Shepherd \& Kurtz
(1999) to within the uncertainties.  A spectral energy distribution
(SED) of G192.16 (Fig. 7) shows our data in relation to previous
measurements.  The fit to the free-free emission at centimeter
wavelengths has a slope of $0.3 \pm 0.07$ (i.e., $S_\nu \propto
\nu^{0.3}$), consistent with a moderately optically thick \HII\
region.  No free-free emission was detected toward G192\,S3.

\subsection{Water maser emission}

VLA observations of G192.16 show a field rich in {\water} maser
emission.  Red-shifted masers located $0.1''$ to $0.2''$ (2000 --
4000\,AU) E-NE of G192.16 have velocities between 6 and 9{\kms}
relative to $v_{LSR}$.  Red and blue-shifted masers are also located
along a stripe to the north of G192.16 where Shepherd \& Kurtz (1999)
discovered an ionized jet of mixed synchrotron and thermal emission.
The masers associated with this jet tend to have red-shifted
velocities to the NE and blue-shifted velocities to the SW along the
jet axis.  Figure 1 shows the locations and velocities of the {\water}
maser emission detected in our observations as well as those detected
previously.  

VLBA {\water} maser images were aligned with the VLA maser images as
described in Section 2.2.  The maser emission is variable in position,
intensity, and velocity although masers exist in every epoch.  VLBA
{\water} masers are detected near the end-points of the
synchrotron-thermal jet in each of the three epochs (Fig. 2).  Masers
associated with the jet tend to be red-shifted to the NE and
blue-shifted to the SW and have similar velocities to those observed
with the VLA.  During the first epoch of VLBA observations, one
position with red and blue-shifted maser emission was detected
100\,mas (200\,AU) NE of G192.16.  

Although every VLA observation (this work and previous observations)
detected {\water} masers near the massive protostar, two of three VLBA
epochs did not detect emission in this region.  {\water} masers are
notoriously variable and those in G192.16 are no exception.  Thus, it
could be that the region near G192.16 did not have masers during our
last two VLBA observations.  It seems unlikely that the VLBA would
resolve out the emission given that the $\sim 1$\,mas resolution of
the VLBA observations corresponds to a size-scale of about 2\,AU at a
distance of 2\,kpc which is larger than typical {\water} maser
features found toward other massive star formation regions.  However,
we cannot rule out this possibility.

\subsection{Imaging at 850\,$\mu$m}

The 850\,$\mu$m image (Fig. 3) has a strong peak (S/N $\sim$ 96)
centered on the G192.16 UC\,\HII\ region (shown as a cross in Fig. 3)
along with an extension to the south where G192\,S3 is detected in
{\nht} emission (shown as a plus symbol in Fig. 3).  This extension is
visible in each of the three individual jiggle maps and is thus likely
to be real.  A similar extension of the emission from G192.16 to
G192\,S3 can be seen in the 1.3\,mm continuum image of Beuther et
al. (2002) and in the 21.3\,$\mu$m image taken by the Midcourse Space
Experiment (MSX, figure not shown) although both data show this
feature only at the $2-3 \sigma$ level.  The 10\% contour level in
Fig. 3 extends nearly an arcminute south of G192.16 suggesting the
presence of molecular gas up to $30''$ beyond G192\,S3.

The mass of G192\,S3 may be determined from the flux in a small
aperture (15$''$) centered on the NH$_3$ peak position; however, this
flux will be contaminated by the extended emission associated with the
massive molecular core surrounding G192.16.  To minimize this affect,
we constructed a model image of the G192.16 core from an azimuthally
averaged radial intensity profile.  In creating this average intensity
profile a 60{\deg} wide sector centered on G192\,S3 was omitted to
ensure that emission from G192\,S3 was not included in the model.  The
resulting model was subtracted from the 850\,$\mu$m image to produce
the image shown in the right panel of Fig. 3.  The flux
density of G192\,S3 in a $15''$ aperture measured in this
model-subtracted image is $130 \pm 45${\mjyb}.  Assuming: (1) a
distance of 2 kpc; (2) a 1:100 dust to gas mass ratio; (3) a total gas
opacity at 850\,$\mu$m of 0.034 cm$^2$\,g$^{-1}$ appropriate for
coagulated dust grains with thin ice mantles (Table 1, column 5 of
Ossenkopf \& Henning 1994); and (4) an average dust temperature of
30K; then the total gas+dust mass of the southern core is $1.3$\Msun.
This mass is a factor of 2 lower than the mass determined from NH$_3$
observations.  The Ossenkopf \& Henning opacities provide a good fit
to the observed SEDs toward star-forming cores (e.g., Evans et
al. 2001; Shirley, Evans, \& Rawlings 2002; Young et al. 2003) but are
probably uncertain to within a factor of a few.  Also, the systematic
uncertainty in aperture photometry performed on the subtracted model
is probably $\sim 50$\% .  Given these uncertainties, the mass
estimates based on {\nht} and 850\,$\mu$m imaging agree remarkably
well and suggests that they are sampling similar material.

The peak flux density of the G192.16 core is $2.10 \pm 0.63${\jyb} and
the integrated flux density is $7.4 \pm 2.3$\,Jy in a 120$''$ aperture
(uncertainties include contributions from systematic flux
calibration).  The derived flux density agrees to within the errors
with Williams et al. (2004).  The Williams et al. flux densities at
450 and 850\,$\mu$m and our 850\,$\mu$m point are shown in the SED in
Fig 7.  Using the same assumptions given above, the total mass is
75{\Msun}.  This is significantly larger than the mass estimate based
on {\nht} emission (0.9\Msun) and is consistent with the
interpretation that the {\nht} emission results from only a small
fraction of molecular material that remains in dense, compact
structures.  In contrast, the 850\,$\mu$m mass estimate of 75{\Msun}
agrees to within the errors with the 117{\Msun} derived from {\ceof}
(Shepherd \& Kurtz 1999).


\section{Discussion}

\subsection{Water maser activity}

It is well known that {\water} masers are commonly observed in both
low- and high-mass star forming regions and the presence of {\water}
masers has been linked to the evolutionary status of star formation in
a region (e.g., Tofani et al. 1995, Kurtz et al. 2000, Codella et
al. 2004).  Most often, masers are associated with jets and outflows
(e.g., Claussen et al. 1998; Torelles et al. 2001, 2003; Moscadelli,
Cesaroni, \& Rioja 2000; Furuya et al. 2003, to name a few) although
there is some evidence that {\water} masers may trace circumstellar
disks around massive protostars (e.g., Cesaroni 1990; Torrelles et
al. 1997; Molinari et al. 2000; Seth et al. 2002; Goddi et al. 2004).

Water masers north of G192.16 appear to trace a collimated jet although
the actual driving source of the jet has not been detected.  The
velocity distribution of masers along the jet axis (e.g., mostly
red-shifted to the NE and mostly blue-shifted to the SW) suggests that
the driving object lies near the center of the masers.

Shepherd \& Kurtz (1999) proposed that the {\water} masers near
G192.16 arose in a 1000\,AU flattened, rotating disk surrounding the
protostar.  The velocities were consistent with Keplerian rotation and
with the rotation axis of a {\ceo} core.  Their proposed 1000\,AU disk
is shown as a grey torus in Figs. 1 \& 2.  Our new observations of
{\water} masers with the VLA and VLBA are consistent with this
interpretation however they do not shed a significant amount of light
on the issue given that the VLBA observations only detected a single
maser spot with both red and blue-shifted emission peaks within the
proposed disk and the VLA observations only detected a hand-full of
red-shifted masers E-NE of G192.16.  Thus, we can neither confirm nor
deny the proposal that a 1000\,AU disk surrounds the G192.16
protostar.

\subsection{Molecular core stability}

Two ammonia cores with very different characteristics are detected
toward the G192.16 star forming region.  The core surrounding the
G192.16 massive protostar is clumpy, roughly 0.15\,pc in diameter
(assuming a distance of 2\,kpc), and has a mass of about 0.9\,{\Msun}.
The {\nht} line is optically thin with a full width at half maximum
(FWHM) of 1.5{\kms} and the velocity field is somewhat chaotic.  The
mass estimate based on 850\,$\mu$m emission from warm dust is
75{\Msun} and compares reasonably well with estimates based on {\ceo}
observations (117\Msun).  The large discrepancy between the mass
derived from {\nht} and masses derived from 850\,$\mu$m and {\ceo}
suggests that more than 99\% of the mass of the cloud resides in
material with a density less than $\sim 10^4$\,cm$^{-3}$.

In contrast, G192\,S3 is compact (diameter = 0.05\,pc), optically
thick, and has a mass of 2.6\,{\Msun}.  The line FWHM is roughly half
that of the G192.16 core (0.77\kms); no velocity gradient is seen
across the core although a gradient could easily be missed
with the 0.31{\kms} resolution of our observations.  The similar mass
estimates from {\nht} (2.6\Msun) and 850\,$\mu$m emission (1.3\Msun)
suggests that both methods may probe the same region.  

The gravitational binding energy of a cloud core is given by $G
M_{cloud}^2/{c_1 R_{cloud}}$ where $M_{cloud}$ is the mass of the
cloud, $R_{cloud}$ is the radius, and $c_1$ is a constant that depends
on the mass distribution ($c_1 = 1$ for $\rho \propto r^{-2}$).  This
can be compared to the cloud turbulent energy to determine if the
cloud is likely to collapse or disperse.  Following Myers (1983), the
one-dimensional turbulent velocity dispersion $\sigma$ is related to
the line width $\Delta v$ by
\begin{equation}
\sigma^2 = \frac{\Delta v^2}{8 ln(2)} - \frac{kT}{m}
\end{equation}
where $m$ is the {\nht} molecule mass (17\,amu$\times m_H$ where $m_H$
is the mass of the hydrogen atom) and we take T = 30\,K.  The cloud
turbulent energy is then $\frac{3}{2} M_{cloud}~\sigma^2$.  A
comparison of the gravitational binding energy and the turbulent
energy of the G192.16 and G192\,S3 cores is given in Table 4.  The
ratio of $E_{turb}/E_{grav}$ is 11.7 for G192.16 and 0.3 for G192\,S3
suggesting that G192\,S3 is gravitationally bound and will eventually
collapse to form a protostar while the {\nht} gas surrounding G192.16
is gravitationally unstable and will likely disperse.  Although there
are numerous uncertainties associated with these estimates (see
previous section), the values are consistent with the scenario that
the G192.16 protostellar outflow and ionizing wind may have
significantly disrupted the natal molecular cloud and are in the
process of dispersing the remaining dense molecular gas.  G192\,S3,
$30''$ (0.3\,pc) south of G192.16, appears to be far enough away from
the disrupting influence of the massive protostar to allow the core to
remain gravitationally bound.

\begin{table}[h!]
\caption{Cloud stability comparison}
\small
\begin{tabular}{|l|c|c|}
\hline
            & G192.16            & G192\,S3\\

\hline
\hline
~~$M_{cloud}$ (\Msun)~~~~ 
                    & 0.9                  & 2.6  \\
~~$R_{cloud}$ (pc)    & $\sim 0.07$          & 0.02 \\
~~~$\rho^\dagger$ (g~cm$^{-3}$)
                    & ~~$3.7 \times 10^{-20}$~~
                    & ~~$3.3 \times 10^{-19}$~~\\ 
~~$n^\dagger$ (cm$^{-3}$)     
                    & $8.2 \times 10^{3}$  & $7.3 \times 10^{4}$\\ 
 & & \\
~~$E_{grav}$ (ergs)   & $8.9 \times 10^{41}$ & $2.4 \times 10^{43}$\\ 
~~$E_{turb}$ (ergs)   & $1.0 \times 10^{43}$ & $7.2 \times 10^{42}$\\ 
~~$E_{turb}/E_{grav}$ & 11.7                 & 0.3 \\
 & & \\
~~$M_J$ (\Msun)       & 0.9                  & 0.3 \\
~~$M_{cloud}/M_J$     & 1                    & 9 \\
\hline
\end{tabular}

\vspace{2mm}
$\dagger$~~Average density assuming a uniform sphere.  
\end{table}
\normalsize

Another method to estimate the cloud stability is to calculate the
Jeans mass (the minimum mass for which a cloud will be bound).  The
Jeans mass is given by $M_J = \frac{4}{3} \pi R_{J}^3 \rho$ where
$R_{J}$ is the Jeans length, $R_{J} \sim \left( \frac{k T}{G m \rho}
\right)$ which we take to be the measured radius of the cloud,
$R_{cloud}$.  To estimate the mass density, $\rho$, we assume a
uniform spherical distribution.  This is clearly not a realistic
assumption however the values derived for the number density, $n$,
based on this assumption (Table 4) are within a factor of two of the
number density calculated based on {\nht} emission (Table 3).  Thus,
we assume $\rho$ probably is good to within a factor of 2 or 3.  The
resulting Jeans mass is given in Table 4.  $M_{cloud}/M_J$ is unity
for the {\nht} gas around G192.16 suggesting that is on the verge of
stability while the ratio is 9 for G192\,S3, consistent with the
interpretation that the cloud has adequate mass to collapse under its
own gravitational potential.

Despite the compact nature of G192\,S3 and indications that it may be
gravitationally unstable, no free-free continuum emission from ionized
gas has been detected at centimeter wavelengths, no {\water} maser
emission has been detected, and no high-velocity molecular gas has
been traced to the core which would indicate the presence of a
molecular outflow.  The detection of 850\,$\mu$m emission toward
G192\,S3 is consistent with the presence of a warm ($\sim 20-30$\,K),
dense molecular core that could be heated by gravitational collapse or
an embedded protostar.  Diffuse 21\,$\mu$m emission is located up to
about $45''$ south of G192.16 in images from the Midcourse Space
Experiment (MSX).  The emission is at a level of $2-3 \sigma$ and
there is no strong peak corresponding with G192\,S3.  The MSX did not
detect emission near G192\,S3 at 12 \& 14.6\,$\mu$m.  

G192\,S3 has been detected in 1.2\,mm continuum emission suggesting
that dust is being heated by an embedded source (Beuther et al. 2002).
Figure 8 presents data from Beuther et al. toward G192.16
(05553+1631).  Contours have been chosen to illustrate that the
G192\,S3 source is detected at a level of $6 \sigma$ and is clearly
separated from G192.16.  The peak emission at the location of G192\,S3
is $\sim 70${\mjyb} in an $11''$ beam.  Following the method of
Hildebrand (1983), the mass of gas and dust is estimated from the
1.2\,mm continuum emission using ${\rm M}_{gas + dust} = \frac{{\rm
F}_{\nu}~ {\rm D}^2} {{\rm B}_{\nu}({\rm T}_d)~ \kappa_{\nu}}$ where D
is the distance to the source, ${\rm F}_{\nu}$ is the continuum flux
density due to thermal dust emission at frequency $\nu$, and ${\rm
B}_{\nu}$ is the Planck function at temperature T$_d$.  We take
$\kappa_{\nu} = 0.006(\frac{\nu}{245 {\rm
GHz}})^{\beta}$~cm$^2$~g$^{-1}$ (Kramer et al. 1998; Shepherd \&
Watson 2002). The opacity index $\beta = 1.5$ appears to be
appropriate between wavelengths of 650~microns and 2.7~mm for
sub-micron to millimeter-sized grains expected in warm molecular
clouds and young disks (Pollack et al. 1994).  We assume the dust
emission is optically thin, the temperature of the dust can be
characterized by a single value, and the gas-to-dust ratio is 100.
Using values of $T_d = 30$~K and $\beta = 1.5$, we find the mass of
gas and dust associated with the 1.2\,mm continuum emission peak of
G192\,S3 is approximately 4~M{\sun}.  This value should be considered
an upper limit since the G192\,S3 peak flux density is probably
contaminated by the bright, extended emission from the early B
protostar, G192.16.  Never-the-less, this value is within a factor of
a few of estimates based on {\nht} and 850\,$\mu$m emission.

Additional 1.2\,mm emission is also detected nearly an arcminute to
the SE of the G192.16 protostar.  This is consistent with the presence
of 21\,$\mu$m and 850\,$\mu$m emission along a ridge to the SE of
G192.16, within which G192\,S3 appears to be embedded.
 

We can compare the properties of G192\,S3 with the well-studied
pre-stellar core L1544 in the Taurus molecular cloud (D = 140\,pc)
that appears to be on the verge of gravitational instability
(e.g., Tafalla et al. 1998; Cioleck \& Basu 2000; Caselli et al. 2002a,
2002b).  The L1544 core has a strong 1.3\,mm continuum source with a
flux density of 2.3\,Jy (90{\mjyb} peak with a $13''$ beam or
225{\mjyb} with a $22''$ beam) corresponding to a gas+dust mass of
3.2{\Msun} with a density of $1.5 \times 10^6$\,cm$^{-3}$
(Ward-Thompson, Motte,\& Andre 1999; Caselli et al. 2002a).  The
N$_2$H$^+$ line width is $\sim 0.3${\kms}.  Based on models of the
chemical evolution in molecular cores, Bergin \& Langer (1997) find
that {\nht} and {N$_2$H$^+$} show little depletion in the central,
high-density environment of a pre-protostellar core because of the low
binding energy of nitrogen.  Thus, both make excellent tracers of very
dense regions where depletion is occurring and are expected to have
similar line widths.  Indeed, observations toward several low-mass
starless cores in {\nht} and {N$_2$H$^+$} show similar line
widths ($\sim 0.2${\kms}) for both molecules (Tafalla et al. 2004).
Assuming {\nht} and {N$_2$H$^+$} trace the same regions, one would
then expect a similar line width in {\nht} as seen in {N$_2$H$^+$},
e.g., $\sim 0.3${\kms} for L1544.

Although the mass of warm gas and dust in both L1544 and G192\,S3 is
similar (a few {\Msun}), G192\,S3 has an {\nht} line width that is
more than double that expected in L1544 while the number density is a
factor of 10 lower.  Massive star formation regions are often
dynamically active, producing inherently broad line profiles because
of large velocity dispersions.  Despite the increased line width,
G192\,S3 still appears to be gravitationally unstable and may be in
the process of collapse.

Could G192\,S3 harbor an L1544-like millimeter continuum peak and not
be detected at 2.7\,mm?  Shepherd et al. (1998) observed the region in
which G192\,S3 is located at 112\,GHz with $\sim 5''$ resolution and
obtained an RMS of 2.1{\mjyb}.  Scaling the 240\,GHz L1544 emission to
112.3\,GHz at a distance of 2\,kpc ($S_\nu \propto \nu^2/D^2$), and
assuming that roughly 40\% of the flux density would be recovered in
the $5''$ beam, we would expect a peak flux density of about 1{\mjyb}
-- below the $1 \sigma$ RMS level in the map.  Thus, G192\,S3 could
contain a millimeter continuum peak like that seen in the low-mass
core L1544 and not be detected at 2.7\,mm.  On the other hand, the
peak flux density in 1.2\,mm continuum emission expected from L1544 at
a distance of 2\,kpc is 10{\mjyb} in an $11''$ beam.  Beuther et
al. detect a peak flux density of about 70{\mjyb} toward G192\,S3.
Thus, it appears that the embedded source in G192\,S3 is probably
somewhat more luminous than L1544.

This brief comparison illustrates that G192\,S3 is not likely to
harbor a massive protostar.  The properties are similar to those
expected for low-mass pre-stellar or protostellar cores.  Follow-up,
sensitive millimeter continuum observations are necessary to verify
this interpretation.  In addition, N$_2$H$^+$ and HCO$^+$ observations
could provide clues as to the evolutionary stage of the cloud collapse
(e.g., Lee et al. 2003).

\section{Conclusions}

We have observed the massive star forming region associated with
G192.16--3.84 in {\nht}(1,1), {\water} masers, and 1.3\,cm continuum
emission, and obtained archive data at 850\,$\mu$m.  The dense gas
associated with G192.16 is clumpy, optically thin, and has a mass of
less than 1\,{\Msun}.  The molecular core appears to be in the process
of being dispersed, perhaps by the molecular outflow and stellar winds
from G192.16.  South of G192.16 is a compact, optically thick {\nht}
core (G192\,S3) with a mass of 2.6\,{\Msun}.  The dense gas is likely
to be gravitationally bound and may be the next site of star formation
in this region.  Based on the presence of 850\,$\mu$m and 1.2\,mm
continuum emission, G192\,S3 may harbor a very young, low-mass
protostar.

Water masers trace an ionized jet $0.8''$ north of G192.16.  Masers
are also located within 500\,AU of G192.16, their velocity
distribution is consistent with the interpretation that the maser
emission arises in a rotating disk however the current observations
do not shed significant light on this issue.

\vspace{4mm} 
\noindent
{\bf Acknowledgments:} D. Shepherd thanks Henrik Beuther for useful
discussions and for providing the fits image of G192.16 at 1.2\,mm.  

T. Borders acknowledges support from the National Science Foundation's
Research Experience for Undergraduates program.

This research made use of data products from the Midcourse Space 
Experiment (MSX).  Processing of the data was funded by the Ballistic 
Missile Defense Organization with additional support from NASA 
Office of Space Science.  This research has also made use of the 
NASA/ IPAC Infrared Science Archive, which is operated by the 
Jet Propulsion Laboratory, California Institute of Technology, 
under contract with the National Aeronautics and Space 
Administration.

This research made use of data from the Canadian Astronomy Data Centre
which is operated by the Dominion Astrophysical Observatory for the
National Research Council of Canada's Herzberg Institute of
Astrophysics.

{}

\clearpage

\noindent
{\bf Figure Captions:}

\noindent {\bf Figure 1:} VLA {\nht}(1,1) line, 1.3\,cm continuum, \&
  {\water} maser emission.  {\bf Top left panel} shows integrated
  emission (moment 0) across the main {\nht} line from 7.24 to
  4.46\kms ($v_{LSR} = 5.7$\kms).  {\nht} is plotted with thin
  contours at --2, 2, 3, 4, 5, 6$\sigma$ (RMS = 2.93\mjybkms) and in
  greyscale from 1$\sigma$ to 17.04\mjyb.  The 1.3\,cm continuum
  (thick contours) image has an RMS of 0.4{\mjyb} and peak 2.8{\mjyb}.
  Contours are plotted at 3, 4, 5$\sigma$ and continue with a spacing
  of $1 \sigma$.  The synthesized beam of $2.9'' \times 2.5''$ at
  P.A. 44.8{\deg} is shown in the bottom right.
  The {\bf top right panel} zooms in on G192.16; colored symbols
  represent {\water} masers detected by the VLA near G192.16 (shown as
  a filled star symbol).  No other {\water} masers were detected
  outside of this region.   
  The {\bf lower right panel} zooms in on a $1.5''$ region around the
  massive protostar.  {\water} masers detected by the VLA in this work
  are shown as small, filled circles with error bars.  Open circles
  represent {\water} masers detected by Shepherd \& Kurtz (1999).
  Shown for reference is a grey torus representing the 1000~AU disk
  suggested by Shepherd \& Kurtz based on their {\water} maser
  observations.  The small grey ellipse at the center of the torus
  marks the position of the B2 protostar G192.16 and the 130\,AU
  accretion disk proposed by SCK01 based
  on 40\,mas resolution observations in 7\,mm continuum emission.  The
  red and blue arrows mark the direction of the large scale molecular
  outflow (Shepherd et al. 1998).  Finally, the black, double arrow
  north of the torus represents the location of the mixed synchrotron
  and thermal jet detected by Shepherd \& Kurtz (1999).
  Colors of the maser symbols represent velocity as shown in the
  position-velocity plot in the {\bf lower left panel}.  The
  dash-dotted vertical line represents $v_{LSR} = 5.7$\kms.

\noindent {\bf Figure 2:} The {\bf top left panel} shows all three
  epochs of VLBA observations of {\water} masers (Epoch 1 = triangles,
  Epoch 2 = squares, Epoch 3 = circles).  {\bf Top right} and {\bf
  bottom left} panels zoom in on the SW and NE maser clumps.  The
  reference position for which all epochs were aligned is shown at
  postion (0,0).  The VLBA masers were aligned to the VLA masers as
  described in the text.  Proposed structures in the top left panel
  are the same as in Fig. 1.  The {\bf bottom right panel} presents a
  position velocity diagram of the masers.

\noindent {\bf Figure 3:} {\bf Left:} SCUBA 850$\mu$m image of
  G192.16.  The UC\,\HII\ region associated with the B2 protostar is
  shown as a cross, the southern core, G192\,S3, is shown as a plus
  symbol.  The center of the image is located at \ra{05}{58}{13.87},
  \dec{+16}{32}{00.1} (J2000).  The image RMS is $\sim 20${\mjyb}.
  Contours are plotted as percentages of the peak flux density of
  2.10{\jyb} (e.g., 10\% (10$\sigma$) means the contours are 10\%\ the
  peak flux density and the lowest contour is at 10$\sigma$).  The
  beam size of $15''$ is shown in the lower left of the image.  {\bf
  Right:} An image of G192\,S3 after an azimuthally averaged radial
  intensity profile of G192.16 has been subtracted (see text for
  details).  Contours begin at $2\sigma$ (40{\mjyb}) and continue at
  spacings of 20\% of the peak flux density (130\mjyb).  Compact knots
  that are roughly the size of the beam at the edge of the
  model-subtracted image are probably due to low-level artifacts in
  the original image.  G192\,S3 is extended and well within the
  $120''$ throw, thus, it is likely to be real.  Symbols are the same
  as in the left panel.

\noindent {\bf Figure 4:} {\nht} channel maps.  Contours = 2, 3, 4, 5,
  6$\sigma$ (RMS = 2.2{\mjyb}).  Velocity in {\kms} is indicated in
  the top left of each panel.  The synthesized beam of $2.93'' \times
  2.55''$ at P.A. 44.8{\deg} and a scale bar of 0.2\,pc are shown in
  the lower right panel.  Top left panel shows the zeroth moment map
  from Fig 1.  The filled triangle indicates the position of the
  G192.16 UC\,\HII\ region.

\noindent {\bf Figure 5:} NH$_3(1,1)$ spectra toward G192.16 (top) and
  G192\,S3(bottom).  Main line peak = 5.7\kms\ for G192.16 and
  6.7\kms\ for the southern core.  The noise is higher for G192\,S3
  because the spectrum was obtained over a smaller area than for
  G192.16.  Thick vertical lines below the spectra mark the locations
  of the main {\nht} transition and the first two hyperfine
  components.

\noindent {\bf Figure 6:} NH$_3(1,1)$ first moment map. Color scale is
  displayed from dark blue: 4.2{\kms} to dark red: 7.2{\kms}.  

\noindent {\bf Figure 7:} The Spectral Energy Distribution (SED) of
  G192.16.  The filled triangle represents the flux density at 1.3\,cm
  and the filled star is the 850\,$\mu$m flux density from this work.
  Other symbols are from: Hughes \& MacLeod (1993) - filled square
  (6\,cm); Shepherd \& Kurtz (1999) - open square (3.6\,cm); SCK01 -
  open triangle (7\,mm); Beuther et al. (2002) - plus symbol; Shepherd
  et al. (1998) - open diamond (3\,mm); Williams, Fuller, \& Sridharan
  (2004) - open circles (850 \& 450\,$\mu$m); MSX - filled circles
  (12.13, 14.65, \& 21.3\,$\mu$m); IRAS - asterisks (100, 60, 25, \&
  12\,$\mu$m).  IRAS points at 100 and 12\,$\mu$m are upper limits.

\noindent {\bf Figure 8:} 1.2\,mm continuum emission from Beuther et
  al. (2002).  The image RMS = 11{\mjyb}.  Contours = 3, 4, 5, 6, 9,
  12$\sigma$ and continue with a spacing of 3$\sigma$.  The UC\,\HII\
  region associated with the B2 protostar is shown as a cross, the
  southern core, G192\,S3, is shown as a plus symbol.



\begin{thebibliography}{}

\vspace{-1mm}
\bibitem[]{} Archibald, E. N., et al. 2002, MNRAS, 336, 1

\vspace{-1mm}
\bibitem[]{} Bergin, E. A. \& Langer, W. D. 1997, ApJ, 486, 316

\vspace{-1mm}
\bibitem[]{} 
Beuther, H., Schilke, P., Menten, K. M., Motte, F., Sridharan, T. K.,
Wyrowski, F. 2002, ApJ, 566, 945

\vspace{-1mm}
\bibitem[]{} Beuther, H., Schilke, P., \& Stanke, T. 2003, A\&A, 408, 601 

\vspace{-1mm}
\bibitem[]{} Beuther, H. \& Schilke, P. 2004, {\it Science}, 303, 1167

\vspace{-1mm}
\bibitem[]{} Caselli, P., Walmsley, C. M., Zucconi, A., Tafalla, M.,
Dore, L., \& Myers, P. C. 2002, ApJ, 565, 344

\vspace{-1mm}
\bibitem[]{} Caselli, P., Walmsley, C. M., Zucconi, A., Tafalla, M., 
Dore, L., \& Myers, P. C. 2002, ApJ, 565, 331

\vspace{-1mm}
\bibitem[]{} Cesaroni, R. 1990, A\&A, 233, 513

\vspace{-1mm}
\bibitem[]{} Cesaroni, R., Felli, M., Jenness, T., Neri, R., Olmi, L.,
Robberto, M., Testi, L., \& Walmsley, C. M. 1999, A\&A, 345, 949

\vspace{-1mm}
\bibitem[]{} Chini, R., Hoffmeister, V., Kimeswenger, S., Nielbock,
M., N\"urnberger, D., Schmidtobreick, L., \& Sterzik, M. 
2004, {\it Nature}, 492, 155

\vspace{-1mm}
\bibitem[]{} Cioleck, G. E. \& Basu, S. 2000, ApJ, 529, 925

\vspace{-1mm}
\bibitem[]{} Claussen, M. J., Marvel, K. B., Wootten, A., \& Wilking,
B. A. 1998, ApJ, 507, 79 

\vspace{-1mm}
\bibitem[]{} Codella, C., Lorenzani, A., Gallego, A.T., Cesaroni, R., \&
Moscadelli, L.  2004, A\&A, 417, 615

\vspace{-1mm}
\bibitem[]{}
Codella, C., Felli, M., Natale, V. 1996, A\&A, 311, 971

\vspace{-1mm}
\bibitem{} Devine, D., Bally, J., Reipurth, B., Shepherd, D. S., \&
Watson, A. M. 1999, ApJ, 117, 2919

\vspace{-1mm}
\bibitem{} Evans, N. J., II, Rawlings, J. M. C., Shirley, Y. L., \& 
Mundy, L. G. 2001, ApJ, 557, 193

\vspace{-1mm}
\bibitem{} Fendt, C. \& Zinnecker, H. 1998, A\&A, 334, 750

\vspace{-1mm}
\bibitem{} Furuya, R. S., Kitamura, Y., Wootten, A., Claussen, M. J.,
Kawabe, R. 2003, ApJS, 144, 71
\vspace{-1mm}
\bibitem{} Goddi, C., Moscadelli, L., Alef, W., \& Brand, J. 2004,
A\&A, in press

\vspace{-1mm}
\bibitem{} Harju, J., Walmsley, C. M., \& Wouterloot, J. G. A. 1993,
A\&AS, 98, 51

\vspace{-1mm}
\bibitem[]{}
Hildebrand, R. H. 1983, Q.J.R.Astr.Soc, 24, 267

\vspace{-1mm}
\bibitem{} Ho, P. T. P. \& Townes, C. H. 1983, ARA\&A, 21, 239

\vspace{-1mm}
\bibitem{} Holland, W. S., et al. 1999, MNRAS, 303, 659 

\vspace{-1mm}
\bibitem{} Hughes, V. A. \& MacLeod, G. C. 1993, AJ, 105, 1495

\vspace{-1mm}
\bibitem{} Indebetouw, R., Watson, C., Johnson, K. E., Whitney, B., \&
Churchwell, E. 2003, ApJ, 596, L83

\vspace{-1mm}
\bibitem{} Jenness, T., \& Lightfoot, J. F. 1998, in ``Astronomical
Data Analysis Software and Systems VII'', ASP Conf. Ser. 145, eds.
R. Albrecht, R. N. Hook and H. A. Bushouse (San Francisco: ASP), 216

\vspace{-1mm}
\bibitem{} K\"onigl, A. 1999 New Astronomy Reviews, 43, 67

\vspace{-1mm}
\bibitem[]{}
Kramer, C., Alves, J., Lada, C., Lada, E., Sievers, A., Ungerechts,
H., \& Walmsley, M. 1998, A\&A, 329, L33

\vspace{-1mm}
\bibitem{} Kurtz, S., Cesaroni, R., Churchwell, E., Hofner, P., \&
Walmsley, C. M. 2000, in ``Protostars and Planets IV'', ed. V. Mannings,
A. Boss \& S. Russell (Tucson: Univ. of Arizona Press), 299

\vspace{-1mm}
\bibitem{} Laughlin, G. \& Bodenheimer, P. 1994, ApJ, 436, 335 

\vspace{-1mm}
\bibitem{} Lee, J.-E., Evans, N. J., II, Shirley, Y. L., Tatematsu,
K. 2003, ApJ, 583, 789

\vspace{-1mm}
\bibitem{} Molinari, S., Brand, J., Cesaroni, R., \& Palla, F. 1996,
A\&A, 308, 573 

\vspace{-1mm}
\bibitem{} Molinari, S., Brand, J., Cesaroni, R., \& Palla, F. 2000,
A\&A, 355, 617

\vspace{-1mm}
\bibitem[]{} Moscadelli, L., Cesaroni, R., \& Rioja, M. J. 2000, A\&A,
360, 663

\vspace{-1mm}
\bibitem{} Myers, P. C. 1983, ApJ, 270, 105

\vspace{-1mm}
\bibitem{} Ossenkopf, V., \& Henning, T. 1994, A\&A, 291, 943

\vspace{-1mm}
\bibitem{} Ouyed, R. \& Pudritz, R.E. 1999, MNRAS, 309, 2330

\vspace{-1mm}
\bibitem[]{}
Pollack, J. B., Hollenbach, D., Beckwith, S., Simonelli, D. P., Roush,
T. \& Fong, W. 1994, ApJ, 421, 615

\vspace{-1mm}
\bibitem{} 
Seth, A. C., Greenhill, L. J., \& Holder, B. P. 2002, ApJ, 581, 325

\vspace{-1mm}
\bibitem{} Sandell, G., Wright, M., \& Forster, J. R. 2003, ApJL, 590,
  L45 

\vspace{-1mm}
\bibitem{} Shepherd, D. S., Watson A. M., Sargent A. I., \& Churchwell
E. 1998, ApJ, 507, 861

\vspace{-1mm}
\bibitem{} Shepherd, D. S. \& Kurtz, S. E. 1999, ApJ, 523, 690

\vspace{-1mm}
\bibitem{} Shepherd, D. S.,  Yu, K. C., Bally,, J. \& Testi, L. 2000
ApJ, 535, 833 

\vspace{-1mm}
\bibitem{} Shepherd, D. S., Claussen M. J., \& Kurtz, S. E. 2001, {\it
Science}, 292, 1513 (SCK01)

\vspace{-1mm}
\bibitem[]{}
Shepherd, D. S. \& Watson, A. M. 2002, ApJ, 566, 1

\vspace{-1mm}
\bibitem{} Shepherd, D.S. 2003, ASP Conf. Ser. 287; Galactic Star
Formation Across the Stellar Mass Spectrum, ed. J.M. De Buizer \&
N. S. van der Bliek (San Francisco: ASP), 333

\vspace{-1mm}
\bibitem{} Shirley, Y. L., Evans, N. J., II, Rawlings, J. M. C., Gregersen,
E. M. 2000, ApJS, 131, 249

\vspace{-1mm}
\bibitem{} Shirley, Y. L., Evans, N. J., II, Rawlings, J. M. C. 2002,
ApJ, 575, 337

\vspace{-1mm}
\bibitem[]{} Shu, F. H., Najita, J. R., Shang, H., \& Li, Z.-Y. 2000,
in Protostars and Planets IV, ed. V. Mannings, A. P. Boss \&
S. S. Russell (Tucson: University of Arizona Press), 789

\vspace{-1mm}
\bibitem[]{}
Snell, R. L., Dickman, R. L., Huang, Y.-L. 1990, ApJ, 352, 139

\vspace{-1mm}
\bibitem[]{} Tafalla, M., Mardones, D., Myers, P. C., Caselli, P.,
Bachiller, R., \& Benson, P. J. 1998, ApJ, 504, 900

\vspace{-1mm}
\bibitem[]{} Tafalla, M., Myers, P. C., Caselli, P., \& Walmsley, C. M
2004, A\&A, 416, 191

\vspace{-1mm}
\bibitem{} Thompson, R. I. 1984 ApJ, 283, 165

\vspace{-1mm}
\bibitem{} Tofani, G., Felli, M., Taylor, G. B., \& Hunter,
T. R. 1995, A\&AS, 112, 299

\vspace{-1mm}
\bibitem[]{}
Torrelles, J. M., Gomez, J. F., Rodr\'{\i}guez, L. F., Ho, P. T. P.,
Curiel, S., \& Vazquez, R. 1997, ApJ, 489, 744

\vspace{-1mm}
\bibitem{} Torelles, J. M., Patel, N. A., G\'omez, J., Ho, P. T. P., 
Rodr\'{\i}guez, L. F., Anglada, G., Garay, G., Greenhill, L., Curiel, S.,
\& Cant\'o, J. 2001, ApJ, 560, 853

\vspace{-1mm}
\bibitem{} Torelles, J. M., Patel, N. A., Anglada, G., G\'omez, J.,
Ho, P. T. P., Lara, L., Alberdi, A., Cant\'o, J., Curiel, S., Garay,
G., \& Rodr\'{\i}guez, L. F. 2003, ApJL, 598, L115

\vspace{-1mm}
\bibitem[]{} Ward-Thompson, D., Motte, F., \& Andre, P. 1999, 
MNRAS, 305, 143

\vspace{-1mm}
\bibitem{} Williams, S. J., Fuller, G. A., \& Sridharan, T. K. 2004,
  A\&A, 417, 115

\vspace{-1mm}
\bibitem{} Wiseman, J. \& Ho P. T. P 1998, ApJ, 502, 676

\vspace{-1mm}
\bibitem{} Yorke, H. W., Bodenheimer, P., \& Laughlin, G. 1995, ApJ,
  443, 199  

\vspace{-1mm}
\bibitem{} Yorke, H. W. \& Sonnhalter, C. 2002, ApJ, 569, 846

\vspace{-1mm}
\bibitem{} Young, C. H., Shirley, Y. L., Evans, N. J., II, \& 
Rawlings, J. M. C. 2003, ApJS, 145, 111

\vspace{-1mm}
\bibitem{} Zhang, Q., Hunter, T. R., \& Sridharan, T. K. 1998, ApJL,
505, L151 

\vspace{-1mm}
\bibitem{} Zhang, Q., Hunter, T. R., Sridharan, T. K., \& Ho,
  P. T. P. 2002, ApJ, 566, 982

\end{thebibliography}
\end{document}